\documentclass[12pt]{iopart}

\def\rp#1{\left(#1\right)}
\def\kp#1{\left\{#1\right\}}

\def\bra#1{\langle{#1}|}
\def\ket#1{|{#1}\rangle}

\def\ketbra#1#2{|{#1}\rangle\langle{#2}|}

\def\matrix22#1#2#3#4{\left(\begin{array}{cc}{#1} & {#2}\\{#3} & {#4}\end{array}\right)}

\def\ba{\begin{align}}
\def\ea{\end{align}}

\usepackage[sort]{cite}
\usepackage{bbm}
\usepackage{color}
\usepackage{graphicx}

\begin{document}

\title[Free induction decay of single spins in diamond]{Free induction decay of single spins in diamond}

\author{J R Maze$^1$, A Dr\'eau$^2$, V Waselowski$^1$, H Duarte$^1$, J-F Roch$^{3}$ and V Jacques$^2$}

\address{$^1$ Departmento de F\'isica Pontificia Universidad Cat\'olica de Chile, Santiago 7820436, Chile}
\address{$^2$ Laboratoire de Photonique Quantique et Mol\'eculaire, CNRS and ENS Cachan UMR 8537, 94235 Cachan, France}
\address{$^{3}$Laboratoire Aim\'e Cotton, CNRS, Universit\'e Paris-Sud and ENS Cachan, 91405 Orsay, France}
\ead{jmaze@uc.cl}

\begin{abstract}
We study both theoretically and experimentally the free induction decay (FID) of the electron spin associated with a single nitrogen-vacancy defect in high purity diamond, where the main source of decoherence is the hyperfine interaction with a bath of $^{13}$C nuclear spins. In particular, we report a systematic study of the FID signal as a function of the strength of a magnetic field oriented along the symmetry axis of the defect. On average, an increment of the coherence time by a factor of $\sqrt{5/2}$ is observed at high magnetic field in diamond samples with a natural abundance of ${^{13}}$C nuclear spins, in agreement with numerical simulations and theoretical studies. Further theoretical analysis shows that this enhancement is independent of the concentration of nuclear spin impurities. By dividing the nuclear spin bath into shells and cones, we theoretically identify which nuclear spins are responsible for the observed dynamics.
\end{abstract}

\maketitle

\section{Introduction}
Color centers in solids have emerged as good candidates for quantum information processing as they provide optical access on demand to a quantum degree of freedom~\cite{Ladd:Nature2008}. Among them, the nitrogen-vacancy (NV) color center in diamond can be manipulated with full control and has remarkable properties for many applications ranging from high sensitivity and high resolution magnetometry~\cite{Taylor:NatPhys2008,Maze:Nature2008,Balasubramanian:Nature2008,Maletinsky:arXiv2011,Rondin:arXiv2011}, to quantum information processing~\cite{Dutt:Science2007,Neumann:Science2008,Fuchs:NatPhys2010,Robledo:Nature2011,Togan:Nature2010,Maurer:Science2012} and imaging in life science~\cite{McGuinness:NatNano2011,Alhaddad:Small2011,Hall:SciRep2012}. Most of these applications rely on the long coherence time of the NV defect electronic spin, which is mainly limited by magnetic interactions with a bath of paramagnetic impurities inside the diamond matrix. Therefore efforts have concentrated on creating diamond samples with a low concentration of impurities~\cite{Tallaire:DRM2006,Balasubramanian:NatMat2009}, controlling the implantation of single NV defects~\cite{Meijer:APL2005,Rabeau:APL2006,Pezzagna:Small2010,Toyli:NanoLett2010,Spinicelli:NJP2011}, manipulating and controlling the dynamics of their spin bath~\cite{Cappellaro:PRL2009,Lange:Science2010,Ryan:PRL2010,Naydenov:PRB2011,Cappellaro:PRA2012} and understanding the interaction between the central spin and its environment both experimentally and theoretically~\cite{VanOort:CP1990,Childress:Science2006,Maze:PRB2008,Mizuochi:PRB2009,Stanwix:PRB2010,Pu:NatComm2011,Zhao:PRB2012}. 

The decoherence of a central spin in the presence of a spin bath has been addressed using several approaches~\cite{Coish:PRB2005,deSousa:PRB2003,Witzel:PRB2005,Witzel:PRB2006,Cyw:PRB2009,Yao:PRB2006}. Here we study both experimentally and theoretically the free induction decay (FID) of the electronic spin associated with a single NV defect in diamond. In particular, we have performed a statistical study on the coherence time, $T_2^*$, as a function of the strength of a magnetic field oriented along the NV defect symmetry axis. Our results indicate an increment of the coherence time at large magnetic fields in agreement with our numerical simulations and theoretical studies. In addition, we study the coherence time for different concentrations of the spin bath and identify the main features of the central spin dynamics. By dividing the bath into shells and cones, we analyse the contribution to decoherence of each impurity with respect to its position relative to the central spin. These results, which complement recent works reported in Refs~\cite{Zhao:PRB2012,BaoLiu:ScientRep2012}, might be particularly useful for diamond based magnetometry of dc fields\cite{Taylor:NatPhys2008,Dreau:PRB2011}, diamond-based quantum information experiments such as nuclear spin bath control and nuclear-spin based quantum memories studies with isotopically engineered diamond samples~\cite{Maurer:Science2012}. 

The paper is organized as follows. We first introduce in Section~\ref{sec:theo} a theoretical model of a central spin immersed in a spin bath, which allows us to infer the phase memory time $T_{2}^{*}$ of the NV defect electron spin. We then describe in Section~\ref{sec:num} the numerical and experimental methods used to analyze the FID signals recorded from single NV defects. The experimental and numerical results are finally presented in Section~\ref{sec:discuss}, and compared to the theoretical predictions.

\section{Model and methods}
\label{S1}

\subsection{Theoretical model}\label{sec:theo}
\begin{figure}[h!]
\begin{center}
\includegraphics[width=0.7\textwidth]{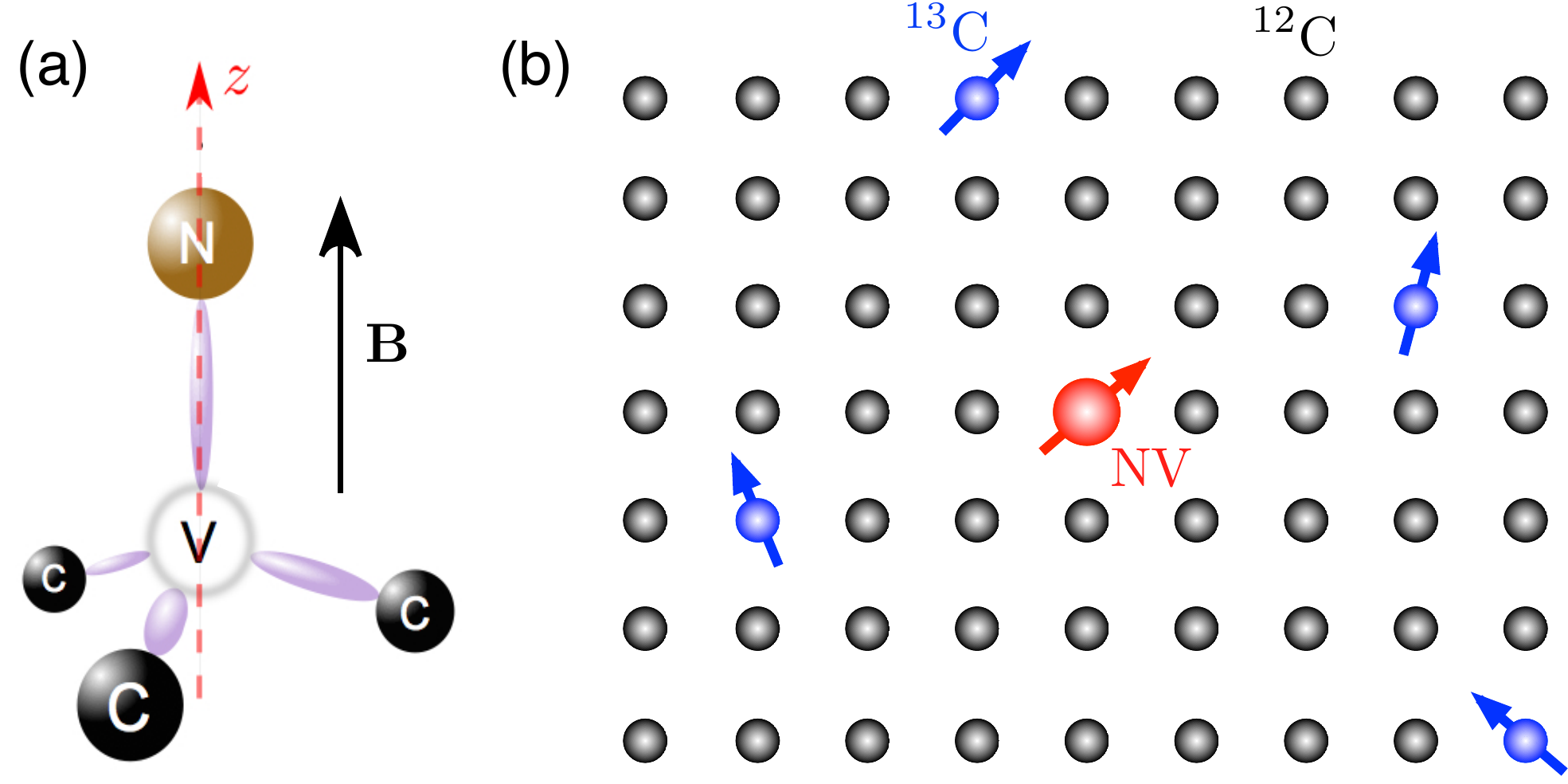}
\caption{(a) Atomic structure of the NV defect in diamond consisting of a substitutional nitrogen atom (N) associated with a vacancy (V) in an adjacent lattice site of the diamond matrix. The NV defect axis $z$ provides an intrinsic quantization axis for the electron spin and a magnetic field $\mathbf{B}$ is applied along this axis. (b) Schematic view of a single NV defect (red arrow) placed in a nuclear spin bath. Blue arrows indicate $^{13}$C nuclear spins located randomly in the diamond lattice. Decoherence of the central electronic spin is induced by hyperfine coupling with the $^{13}$C nuclear spins.}
\label{fig0}
\end{center}
\end{figure} 
We consider the electronic spin $S=1$ associated with a single negatively-charged NV defect in diamond (figure~\ref{fig0}(a)). This central spin is nestled in a high-purity diamond lattice, where electron spin impurities like nitrogen donors (P1 centers) are below $1$ ppb and thus do not contribute to the decoherence of the central spin. Each lattice position can be occupied either by $^{12}$C atoms (spinless) or by $^{13}$C isotopes (nuclear spin $I=\frac{1}{2}$), which form a nuclear-spin bath (figure~\ref{fig0}(b)). The natural abundance of ${^{13}}$C in diamond is $p_{nat}=1.1\%$ while isotopically-modified diamond samples exhibit ${^{13}}$C concentrations as low as $p=0.01\%$~\cite{Ishikawa:NanoLett2012}. In a high purity diamond sample, the decoherence of the NV defect electron spin is dominated by hyperfine coupling with the $^{13}$C nuclear spins. The full Hamiltonian of the system is given by 
\begin{eqnarray}
H = H_{\rm{NV}} + H_{\rm{bath}} + H_{\rm{NV-bath}} \ .
\end{eqnarray}

In a magnetic field $\mathbf{B}$, the Hamiltonian describing the NV defect $H_{\rm{NV}}$, the bath $H_{\rm{bath}}$ and their interaction $H_{\rm{NV-bath}}$ read 
\begin{eqnarray}
H_{\rm{NV}}  =  DS_z^2 + \gamma_e\mathbf{B}\cdot \mathbf{S},\\
H_{\rm{bath}}  =  \sum_n \gamma_n \mathbf{B}\cdot\mathbf{I}^{(n)} + \sum_{n<m} \mathbf{I}^{(n)}\cdot\mathbbm{C}^{(nm)}\cdot\mathbf{I}^{(m)}, \\
H_{\rm{NV-bath}}  =  \sum_n \mathbf{S}\cdot\mathbbm{A}^{(n)}\cdot\mathbf{I}^{(n)} \ ,
\end{eqnarray}
where $D/2\pi=2.87$ GHz is the zero field splitting of the NV defect electron spin, $\gamma_e/2\pi = 2.8$ MHz/G (resp. $\gamma_n/2\pi = 1.07$ kHz/G) is the gyromagnetic ratio of the electronic (resp. nuclear) spin, $\mathbbm{A}^{(n)}$ is the hyperfine tensor between the central spin and nuclear spin $n$, and $\mathbbm{C}^{(nm)}$ is the dipole-dipole interaction between nuclei $n$ and $m$. We note that we consider magnetic field amplitudes large enough to neglect strain-induced splitting terms in equation (2). The decoherence of the NV defect electron spin in the very low magnetic field regime ($B<0.1$~G), for which strain-induced splitting protects the central spin against magnetic field fluctuations, has been addressed elsewhere~\cite{Dolde:NatPhys2011}. 

We focus on the free induction decay (FID) of a coherent superposition of the central spin as a function of the amplitude of a magnetic field applied along the NV defect axis, denoted $z$ (figure~\ref{fig0}(a)). In this condition, the electron spin quantization axis is fixed by the NV defect axis and the eigenstates of $H_{\rm{NV}}$ are denoted as $\ket{m_{s}}$ with $m_{s}=0,\pm1$. The FID signal can be obtained by performing a Ramsey sequence $\frac{\pi}{2}-\tau-\frac{\pi}{2}$ on the central spin, where a first $\frac{\pi}{2}$ pulse rotates the central spin from a prepared state $\ket{0}$ to a coherent superposition $\ket{+}=\frac{1}{\sqrt{2}}\left(\ket{0}+\ket{1}\right)$ which evolves over a time $\tau$. During such a free precession time, the central spin only interacts with the external magnetic field and the nuclear spin bath. Finally, a second $\frac{\pi}{2}$ pulse rotates the central spin again, thus converting residual spin coherence into population. The FID signal can be interpreted as the probability to retrieve the initial state $\ket{0}$ and can be formally calculated as 
\begin{eqnarray}\label{eq:signal}
s(\tau) &=& {\rm Tr}_{bath}\left[\bra{0}\rho(\tau)\ket{0}\right],\\
\rho(\tau) &=& U_T(\tau)\rho_e\otimes\rho_{bath} U_T(\tau)^\dagger \ ,
\end{eqnarray}
where the trace is taken over the Hilbert space of the bath. Here $\rho_e=\ketbra{0}{0}$ is the initial electronic spin density matrix and $\rho_{bath} = \prod_n \rho_n^\otimes$ is the initial density matrix of the nuclear spin bath. The total unitary evolution operator of the system is denoted $U_T(\tau) = R(\frac{\pi}{2})U(\tau)R(\frac{\pi}{2})$, where $R(\frac{\pi}{2})$ corresponds to a $\frac{\pi}{2}$-rotation operator over the electronic spin and $U(\tau)=\exp(-iH\tau)$.

We neglect the slow dynamics related to dipolar interactions between nuclear spins within the bath, which is at most $2$ kHz for nearest neighbor interaction. Indeed, experiments have shown that the NV defect coherence time $T_{2}^{*}$ does not exceed few $\mu$s in diamond samples with natural abundance of ${^{13}}$C~\cite{Mizuochi:PRB2009}, indicating that the interaction between the central spin and the bath induces decoherence at a much faster rate than the dipolar coupling between ${^{13}}$C's within the bath. In addition, the large value of the zero field splitting allows us to perform the secular approximation for the electronic spin even at small magnetic field values. Therefore, the terms proportional to $S_x,S_y$ are neglected in the Hamiltonian. For a magnetic field pointing along the $z$-axis, the Hamiltonian can thus be written as
\begin{eqnarray}\label{eq:h}
H = DS_z^2 + \gamma_e BS_z + \sum_n \gamma_n BI^{(n)}_{z} + \sum_{n} S_z\mathbf{A}^{(n)}\mathbf{I}^{(n)} \ ,
\end{eqnarray}
where the last two terms sum over each nuclear spin $n$ and $\mathbf{A}^{(n)} = (A_{zx}^{(n)}, A_{zy}^{(n)}, A_{zz}^{(n)})$ is now a hyperfine vector. For Ramsey experiments, we focus on the rotating frame at which the electron spin transition $|0\rangle \rightarrow |1\rangle$ is addressed so that the Hamiltonian reduces to the last two terms in equation (\ref{eq:h}). We note that this Hamiltonian is diagonal in the electronic spin subspace and therefore it can be considered as a Hamiltonian for the nuclei, $H_{m_s}$, that is conditional to the state of the electron spin $m_s$. The full Hamiltonian of the system can thus be written as
\begin{eqnarray}
H = \sum_{m_{s}=0}^{1} \ket{m_{s}}\bra{m_{s}}\otimes  \ H_{m_s} \ ,
\end{eqnarray}
\begin{eqnarray}
H_{m_s} = \sum_n H_{m_s}^{(n)}=\sum_n \mathbf{\Omega}_{m_s}^{(n)}\cdot \mathbf{I}^{(n)},
\end{eqnarray}
where $\mathbf{\Omega}_{m_s}^{(n)} = \gamma_n\mathbf{B} + m_s\mathbf{A}^{(n)}$ represents the vector around which the nuclear spin $n$ rotates, {\it i.e.} the Larmor vector. The evolution of the bath can then be written as the direct product of the evolution operator of each nuclear spin, $U_{m_s}(\tau) = \prod_n U_{m_s}^{(n)}(\tau)$, where $U_{m_s}^{(n)}(\tau) = \exp(-iH_{m_s}^{(n)}\tau)$.

It follows from equation (\ref{eq:signal}) that the FID signal is given by
\begin{eqnarray}\label{eq:s}
s(\tau) = \frac{1}{2} - \frac{1}{2}{\rm Re}\prod_n\mathcal{S}_n(\tau) \ ,
\end{eqnarray}
where the product runs over all nuclear spins and $\mathcal{S}_n = {\rm Tr}_n\left[\rho_n U_{1}^{(n)\dagger} U_{0}^{(n)}\right]$ is the contribution of nuclear spin $n$ to the signal. 
In order to obtain the effect of the nuclear spin bath on the FID signal envelope, we exclude the contribution of the intrinsic $^{14}$N nuclear spin of the NV defect ($I=1$) and of strongly coupled nuclear spins because they only contribute to coherent oscillations of the signal~\cite{Childress:Science2006}. For all other nuclear spins, we assume an initial state given by the high temperature limit, $\rho_n= \mathbbm{1}/2 $. In this case,
\begin{eqnarray}\label{eq:psu}
\mathcal{S}_n(\tau)  &=& \cos\frac{\Omega_{0}^{(n)}\tau}{2}\cos\frac{\Omega_{1}^{(n)}\tau}{2} + \cos\beta^{(n)}  \sin\frac{\Omega_{0}^{(n)}\tau}{2}\sin\frac{\Omega_{1}^{(n)}\tau}{2} ,
\end{eqnarray}
where $\beta^{(n)}$ is the angle between $\mathbf{\Omega}_{0}^{(n)}$ and $\mathbf{\Omega}_{1}^{(n)}$ (see Appendix).

As a figure of merit we consider the envelope of the FID signal,
\begin{eqnarray}\label{eq:envelope}
E(\tau) = {\rm Re}\prod_n\mathcal{S}_n(\tau) \ ,
\end{eqnarray}
where the product runs over lattice positions occupied by a ${^{13}}$C atom. For a diamond lattice with a given concentration $p$ of ${^{13}}$C not all positions of the lattice are occupied by a ${^{13}}$C. We therefore define the random variable $x_n = \kp{1,0}$, with mean $p$, to denote the presence ($x_n=1$) or absence ($x_n=0$) of a ${^{13}}$C at the lattice position $n$. For a given configuration of ${^{13}}$C $\kp{x_n}$, the FID envelope can be written as
\begin{eqnarray}\label{eq:envelopec}
E(\tau, \kp{x_n}) = {\rm Re}\prod_n\left[x_nS_n(\tau) + 1-x_n\right] \ .
\end{eqnarray}
We note that the product now runs over all positions of the lattice. Assuming that random variables $x_n$ are independent, the average envelope is,
\begin{eqnarray}\label{eq:envelopep}
\bar{E}(\tau) = {\rm Re}\prod_n\left[pS_n(\tau) + 1-p\right] \ .
\end{eqnarray}
We will use this last equation to analyze the average trend of the FID signal in the next sections. We now briefly discuss two simple limits to gain an insight into the FID dynamics.

In the weak magnetic field limit ($\gamma_n B\ll A^{(n)}$), the FID envelope is given by,
\begin{eqnarray}
\bar{E}(\tau) &\approx& \prod_n\rp{1 - \frac{p}{8}[A^{(n)}\tau]^2} =e^{-(\tau/T_{2,\rm{\tiny{ LF}}}^*)^2}, \label{ELF}\\
T_{2,\rm{\tiny{ LF}}}^* &=& \rp{ \frac{p}{8}\sum_n [A^{(n)}]^2}^{-1/2} \ ,
\label{T2LF}
\end{eqnarray}
where $[A^{(n)}]^2 = [A_{zx}^{(n)}]^2 +[A_{zy}^{(n)}]^2 + [A_{zz}^{(n)}]^2 $. The decay of the FID signal is Gaussian and decoherence is caused because each nuclear spin precesses differently with magnitude $A^{(n)}$. 

In the large magnetic field limit, $\gamma_n B\gg A^{(n)}$, since $\cos\beta^{(n)} \sim 1$, the FID envelope reads
\begin{eqnarray}
\bar{E}(\tau) &\approx& \prod_n\rp{1 - \frac{p}{8}[A_{zz}^{(n)}\tau]^2} = e^{-(\tau/T_{2,\rm{\tiny{ HF}}}^*)^2},\label{EHF} \\
T_{2,\rm{\tiny{ HF}}}^* &=& \rp{ \frac{p}{8}\sum_n [A_{zz}^{(n)}]^2}^{-1/2} \ .
\label{T2HF}
\end{eqnarray}
The signal also exhibits a Gaussian decay and decoherence is only due to the difference in Larmor precession along the direction of the external magnetic field because the anisotropic components of the hyperfine interaction are suppressed by the nuclear Zeeman energy. Since $A_{zz}^{(n)}\leq A^{(n)}$ is always true, the NV defect coherence time is enhanced at high magnetic field. The enhancement factor $\eta$ can be calculated as 
\begin{eqnarray}\label{eq:enhancement}
\eta=\frac{T_{2,\rm{\tiny{ HF}}}^*}{T_{2,\rm{\tiny{ LF}}}^*} =\sqrt{\frac{\sum_n [A^{(n)}]^2}{\sum_n [A_{zz}^{(n)}]^2}}.
\end{eqnarray}
As justified in section~\ref{sec:num}, we neglect the contact interaction and therefore consider a pure dipole-dipole interaction between the central spin and the nuclei. The hyperfine components of nuclear spin $n$ can then be written as
\begin{eqnarray}\label{eq:A}
A^{(n)}_{zz}&=& \alpha_n(1-3\cos^2\theta_n),\\
\label{eq:B}
A^{(n)}_{\perp} &=& \sqrt{\left[A_{zx}^{(n)}\right]^2 + \left[A_{zy}^{(n)}\right]^2} = 3\alpha_n\sin\theta_n\cos\theta_n,\\
A^{(n)} &=& \sqrt{\left[A^{(n)}_{\perp}\right]^2 + \left[A_{zz}^{(n)}\right]^2} = \alpha_n\sqrt{1+3\cos^2\theta_n}.
\end{eqnarray}
where $\alpha_{n}$ is the magnitude of the interaction and $\theta_n$  is the angle between the NV symmetry axis and the imaginary line that joins the central spin and nucleus $n$. Considering a homogeneous distribution of nuclei and integrating over $\theta$ and $\phi$, the average enhancement factor is finally given by
\begin{eqnarray}\label{eq:enhancement}
\eta=\frac{T_{2,\rm{\tiny{ HF}}}^*}{T_{2,\rm{\tiny{ LF}}}^*} =\sqrt{\frac{5}{2}} = 1.58... \ ,
\label{enhance}
\end{eqnarray}
independent of the concentration of nuclear spin impurities.

\subsection{Numerical and experimental methods}\label{sec:num}

The FID signal is numerically simulated using two different approaches. The first method (N1) consists in a statistical analysis over a large number of randomly sorted ${^{13}}$C distributions $\kp{x_n}$ for which equation (\ref{eq:envelopec}) is used to calculate the envelope of the FID signal. For each ${^{13}}$C distribution, the result of the simulation is fitted to the function ${\rm exp}\left[-(\tau/T_2^*)^{\epsilon}\right]$ in order to extract the exponent of the FID signal decay~$\epsilon$ and the coherence time $T_2^*$. The second approach (N2) consists in calculating the average envelope of the FID signal using equation (\ref{eq:envelopep}), thus considering that each lattice site has a probability $p$ to host a $^{13}$C. This approach takes less computational time to infer the mean behavior of the single electronic central spin which otherwise must be obtained using the first approach by averaging the signal for a large number of different ${^{13}}$C distributions. For both methods, a lattice of 100,000 atoms is considered, corresponding to approximatively 1000 nuclear spins for a diamond sample with natural abundance of ${^{13}}$C ($p_{nat}=1.1\%$). In addition, the nuclear spin bath is divided into shells and cones in order to study the contribution to the FID dynamics of each ${^{13}}$C with respect of its position relative to the central spin. 

We only consider the dipolar part of the hyperfine interaction in our model. Although a strong contact interaction has been measured for several ${^{13}}$C's close to the central spin~\cite{Smeltzer:NJP2011,Dreau:2012} and calculated theoretically~\cite{Gali:PRB2009}, we restrict the study to low $^{13}$C concentrations $p\leq 4\%$, where decoherence is dominated by the dipolar interaction with the nuclear spin bath~\cite{Mizuochi:PRB2009}. In this framework, strongly coupled nuclear spins only contribute to coherent oscillations of the FID signal but do not change its envelope. Therefore, we exclude from our simulations the lattice sites associated with nuclei that strongly interact with the central spin such as the well known 130-MHz-splitting linked to the nearest neighbor sites of the vacancy~\cite{Felton:PRB2009} and those with a contact interaction larger than few MHz. The neglected lattice positions correspond to about 100 lattice points, {\it i.e.} on average one nuclear spin for a natural abundance of ${^{13}}$C.

From the experimental side, individual NV defects hosted in a high-purity diamond crystal with a natural abundance of ${^{13}}$C are optically isolated at room temperature using a scanning confocal microscope. A laser operating at $532$ nm wavelength is focused onto the sample through a high numerical aperture oil-immersion microscope objective. The NV defect photoluminescence (PL) is collected by the same objective and directed to a photon counting detection system. Under optical illumination, the NV defect is efficiently polarized into the $m_s=0$ spin sublevel~\cite{Manson:PRB2006}. This process provides an efficient state preparation for the Ramsey sequence. In addition, the PL intensity is significantly higher ($\sim 30 \%$) when the $m_{s}=0$ state is populated. Such a spin-dependent PL response enables the detection of electron spin resonances (ESR) on a single NV defect by optical means~\cite{Gruber:Science1997}. As discussed in the previous section, the FID signal is measured by applying a Ramsey sequence $\frac{\pi}{2}-\tau-\frac{\pi}{2}$ to the NV defect electron spin (figure \ref{fig2}(a)). For that purpose, the ESR transition $|0\rangle \leftrightarrow |1\rangle$ is driven with a microwave field applied through a copper microwire directly spanned on the diamond surface. In addition, a permanent magnet placed on a three-axis translation stage is used to apply a static magnetic field with controlled orientation and amplitude. The magnetic field is aligned along the NV axis with a precision better than $1^{\circ}$.

\section{Results and Discussion}\label{sec:discuss}

\subsection{Enhancement of the coherence time at high magnetic field}

\begin{figure*}
\begin{center}
\includegraphics[width=0.98\textwidth]{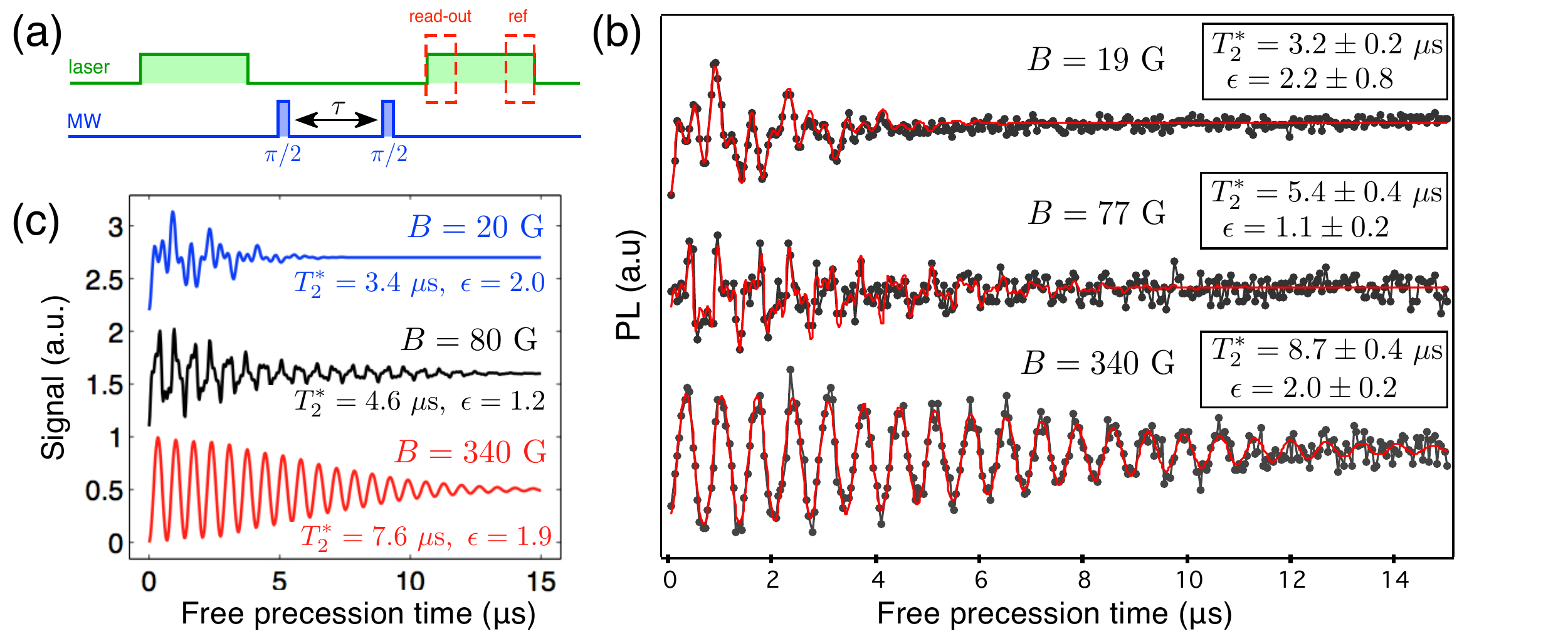}
\caption{(a) Experimental sequence used to measure the FID of the NV defect electron spin. A first laser pulse ($2 \ \mu$s) is used to polarize the electron spin in the $m_{s}=0$ sublevel. A Ramsey sequence consisting of two resonant microwave (MW) $\pi/2$-pulses separated by a variable free evolution duration $\tau$ is then applied and a second laser pulse is finally used for spin state read-out. For data analysis, the NV defect PL recorded during the first $300$~ns of the laser pulses is used for spin-state read-out while the PL recorded during the last $300$~ns is used as reference. (b) Typical FID signals recorded from a single NV defect (NV12) for three different magnetic field amplitudes applied along the NV defect axis. The red solid line is data fitting as explained in the main text. The results of the fitting for the different magnetic fields are as follows. For $B = 19$ G, $\beta_{+1} = 0.347$, $\beta_{0} = 0.333$, $\beta_{-1} = 0.320$ and $\delta = -0.61$ MHz. For $B = 77$ G, $\beta_{+1} = 0.565$, $\beta_{0} = 0.288$, $\beta_{-1} = 0.147$ and $\delta = 3.65$ MHz. And for $B = 340$ G, $\beta_{+1} = 1$, $\beta_{0} = 0$, $\beta_{-1} = 0$ and $\delta = 3.63$ MHz. (c) Numerical simulations of the FID signal for a particular ${^{13}}$C distribution. These graphs are obtained by multiplying the simulated FID envelope by the magnetic field dependent popullations $\beta_{m}$ of each $^{14}$N nuclear spin state.}
\label{fig2}
\end{center}
\end{figure*}

We first study the dynamics of the FID signal as a function of the amplitude of a magnetic field applied along the NV defect axis for a natural abundance of $^{13}$C. Typical FID signals recorded from a single NV defect at different magnetic field magnitudes are shown in figure~\ref{fig2}(b). Since the intrinsic nitrogen atom of the NV defect is a $^{14}$N isotope ($99.6 \%$ abundance), corresponding to a nuclear spin $I=1$, each electron spin state is split into three sublevels by hyperfine interaction. The ESR spectrum thus exhibits three hyperfine lines, splitted by $A_{\rm N}=-2.16$ MHz~\cite{Felton:PRB2009} and corresponding to the three nuclear spin projections. As a result, beating frequencies can be observed in the FID signal at low field (figure~\ref{fig2}(b), upper and middle traces). By increasing the magnetic field, the NV defect gets closer to the excited-state level anti-crossing (LAC) which occurs at $B_{\rm LAC}\approx 510$~G~\cite{Fuchs:PRL2008,Neumann:NJP2009}. In this case, electron-nuclear-spin flip-flops mediated by hyperfine interaction in the excited-state lead to an efficient polarization of the $^{14}$N nuclear spin~\cite{Jacques:PRL2009}. Consequently, the beating frequencies gradually disappear in the FID signal while increasing the magnetic field magnitude and the oscillations are then only due to the detuning of the microwave excitation (figure~\ref{fig2}(b), lower trace).  

From these measurements, the exponent of the FID decay $\epsilon$ and the coherence time $T_2^*$ are extracted through data fitting with the function $\exp[-(\tau/T_{2}^{*})^{\epsilon}]\sum_{m=-1}^{m=1}\beta_{m}\cos[2\pi (\delta+mA_{\rm N}) \tau]$, where $\delta$ is the microwave detuning and $\beta_{m}$ is the population of each $^{14}$N nuclear spin state. As expected from the model developed in the previous section, we observe a significant enhancement of the coherence time with the amplitude of the magnetic field~(figure~\ref{fig2}(b)). In the high field limit, since the nuclear Zeeman energy is much larger than the strength of the hyperfine coupling, the quantization axis is fixed by the external magnetic field for all $^{13}$C nuclear spins and the anisotropic components of the hyperfine interaction $A_{\perp}^{(n)}$ are suppressed (see equation~(\ref{T2HF})). In the weak field limit, the quantization axis of each nuclear spin is rather given by their hyperfine fields and all the components of the hyperfine interaction contribute to decoherence of the central spin (see equation~(\ref{T2LF})). For a specific distribution of $^{13}$C, the numerical simulation reproduces well the behavior of the experimental results, as shown in figure~\ref{fig2}(c).
 \begin{figure*}[t]
\begin{center}
\includegraphics[width=0.8\textwidth]{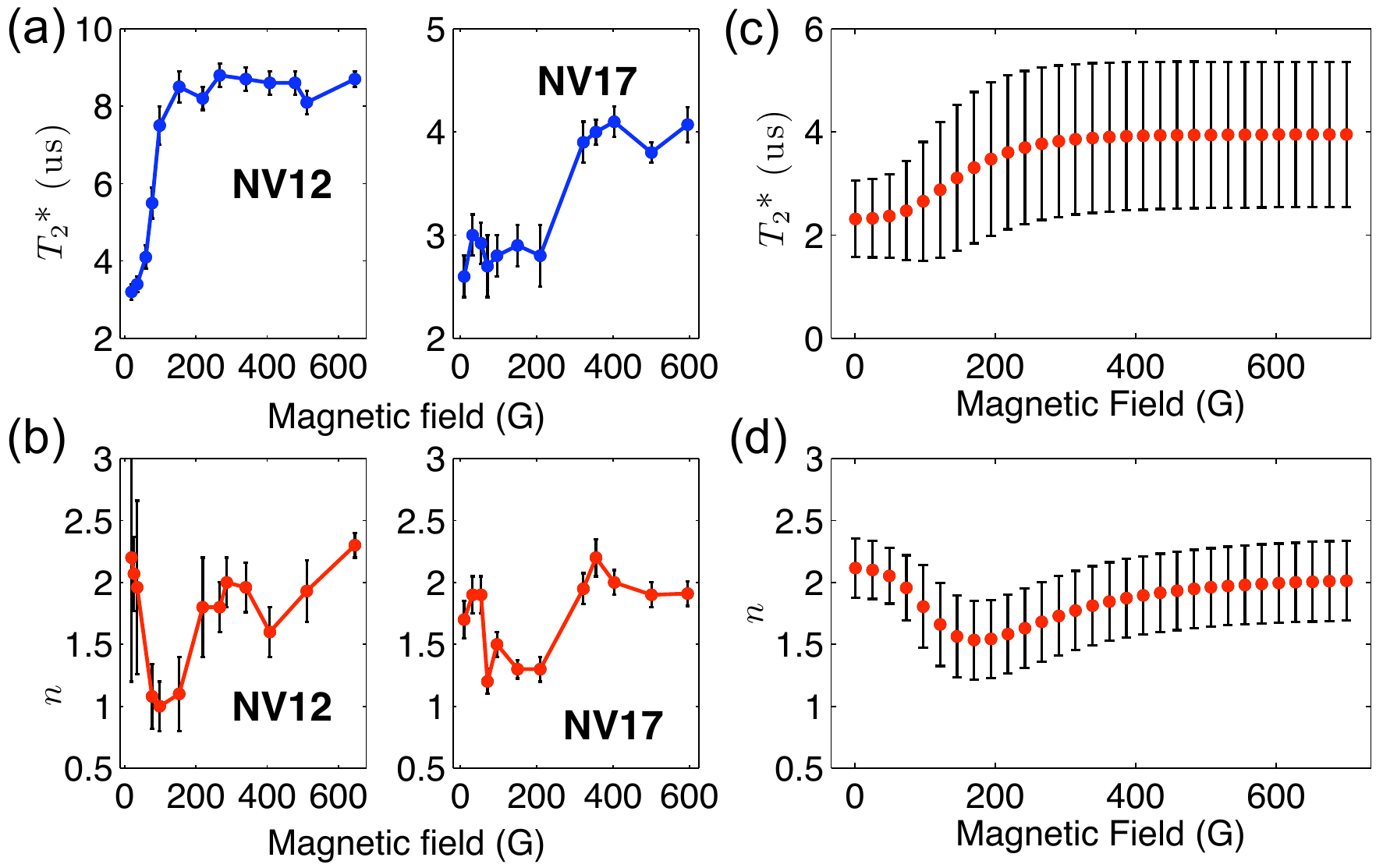}
\caption{(a) Coherence time $T_2^*$ and (b) exponent of the FID decay $\epsilon$ as a function of the magnetic field strength for two single NV defects  (NV12 and NV17). (c) Average coherence time and (d) exponent decay extracted from the numerical simulation of the FID signal over 1000 different nuclear spin bath configurations with $p=1.1\%$. Error bars indicate standard deviation.}
\label{fig3}
\end{center}
\end{figure*} 

To gain more insights into the FID dynamics, the coherence time $T_2^*$ and the exponent of the FID decay $\epsilon$ are plotted in figures~\ref{fig3}(a) and (b) as a function of the magnetic field amplitude for two different NV defects. For both, the coherence time increases with the magnetic field strength. In addition, we also observe that the exponent of the FID decay deviates from $\epsilon=2$ in the magnetic field range where the coherence time is rising, as recently reported in Ref.~\cite{BaoLiu:ScientRep2012}. In this intermediate magnetic field regime, the nuclear Zeeman energy and the hyperfine interaction are of the same order of magnitude. Dephasing of the central spin is due to comparable nuclear rotations along non-parallel precession axis, $\mathbf{\Omega}_{0}$ and $\mathbf{\Omega}_{1}$. A more detailed expression for the contribution from a single nucleus to the signal is given on the Appendix for the intermediate regime. The simulation of the FID signal using the numerical method N1 agrees fairly well with our experimental measurements, as shown in figures~\ref{fig3}(c) and (d). The average coherence time over 1000 different configurations of $^{13}$C is shown on figure~\ref{fig3}(c) as a function of the magnetic field. The coherence time increases around $B=150$~G and at the same time the exponent of the exponential fit deviates from a Gaussian decay ($\epsilon=2$) (Fig.~\ref{fig3} (d)).

 \begin{figure*}[t]
\begin{center}
\includegraphics[width=0.85\textwidth]{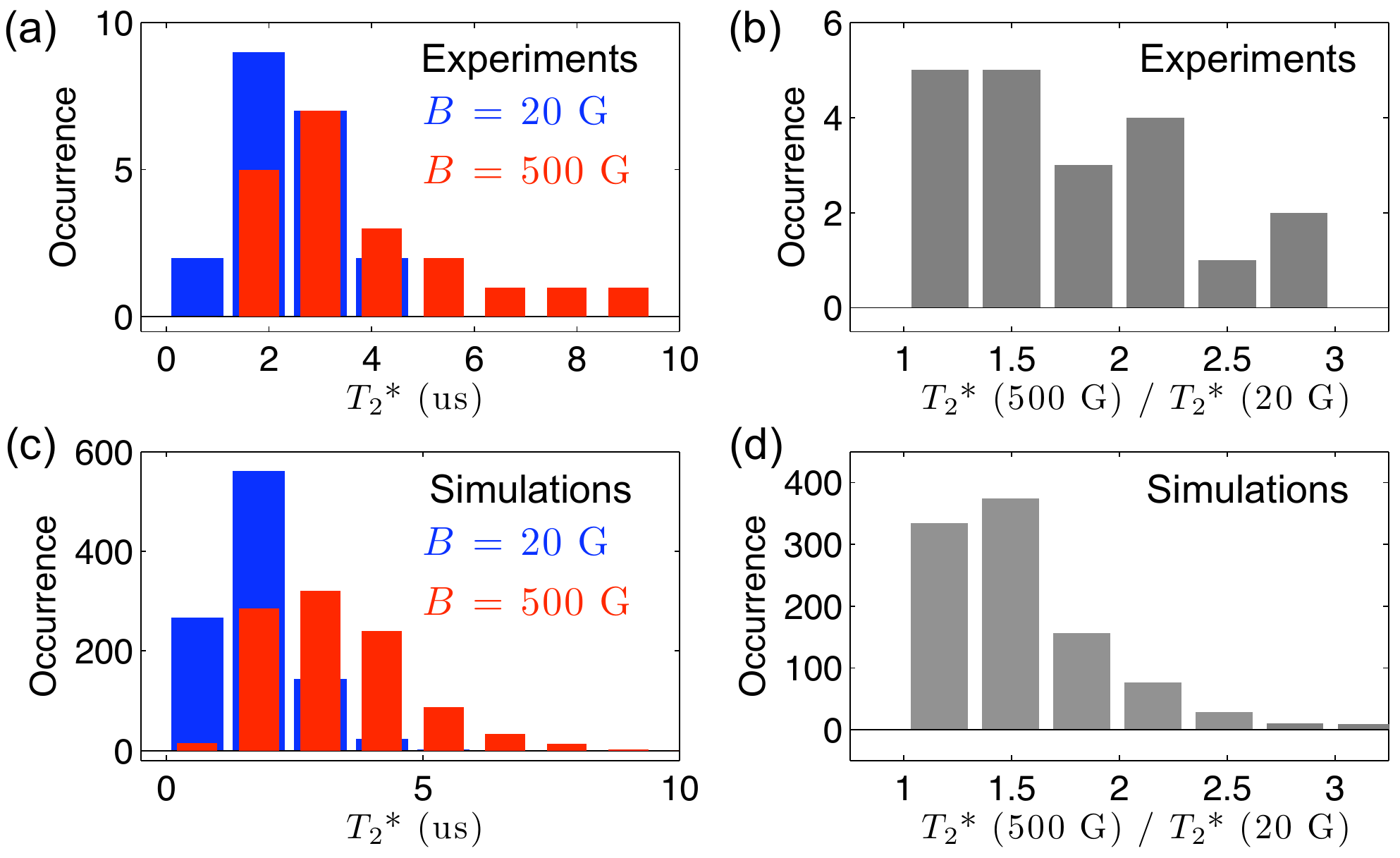}
\caption{(a) Experimental and (c) theoretical histograms of the coherence time $T_2^*$ over 20 and 1000 different ${^{13}}$C bath configurations, respectively. (b) Corresponding experimental and (d) theoretical distributions of the coherence time enhancement between weak magnetic field ($B=20$~G) and high magnetic field ($B=500$~G).}
\label{fig4}
\end{center}
\end{figure*} 

In order to extract a quantitative information of the enhancement factor $\eta$, the coherence time was measured at low field ($B\approx 20$~G) and at high field ($B\approx 500$~G) for a set of 20 single NV defects. Figure \ref{fig4}(a) shows a histogram of the coherence time in both regimes. A large variation of dephasing times is observed owing to the random distribution of $^{13}$C around the central spin. The mean decoherence time at low field is $\langle T_{2}^{*} \, (20{\rm~G})\rangle_{\rm exp}=2.3\pm 1 \ \mu$s while in the high field limit $\langle T_{2}^{*} \, (500{\rm~G})\rangle_{\rm exp}=3.7\pm2 \ \mu$s (errors are standard deviations). The histogram of the enhancement factor $\eta$ is shown in figure \ref{fig4}(b), corresponding to $\langle\eta\rangle=1.7\pm 0.5$, in good quantitative agreement with the model developed in the previous section (see equation~(\ref{enhance})). The histogram of the coherence time extracted from the numerical simulation over 1000 different nuclear spin bath configurations is shown in figure \ref{fig4}(c), leading to $ \langle T_{2}^{*} \, (20{\rm~G})\rangle_{{\rm N1}}= 2.4\pm 0.7 \ \mu$s and $\langle T_{2}^{*} \, (500{\rm~G})\rangle_{\rm N1}= 3.9 \pm 1.4 \ \mu$s, meanwhile the histogram of the enhancement factor is shown in figure \ref{fig4}(d), leading to $\langle\eta\rangle_{{\rm N1}}= 1.7\pm 0.4$, in agreement with the experimental data.

We note that the spread of coherence times increases at high magnetic field indicating that for some particular distributions of the nuclear spin bath there is a sizable enhancement, meanwhile for others there is little enhancement, as shown in figure~\ref{fig3}(a) for two single NV defect having similar coherence times at low magnetic field. This feature is further illustrated in figures \ref{fig4}(c) and (d) which show the histograms of the enhancement factor inferred from the experiments and the numerical simulations, respectively. The contribution to the coherence time enhancement of each ${^{13}}$C with respect to its position relative to the central spin is addressed in Section~\ref{sec:shells} where the bath is divided into shells and cones.

\subsection{Effect of the $^{13}$C nuclear spin concentration}
\begin{figure*}[t]
\begin{center}
\includegraphics[width=0.85\textwidth]{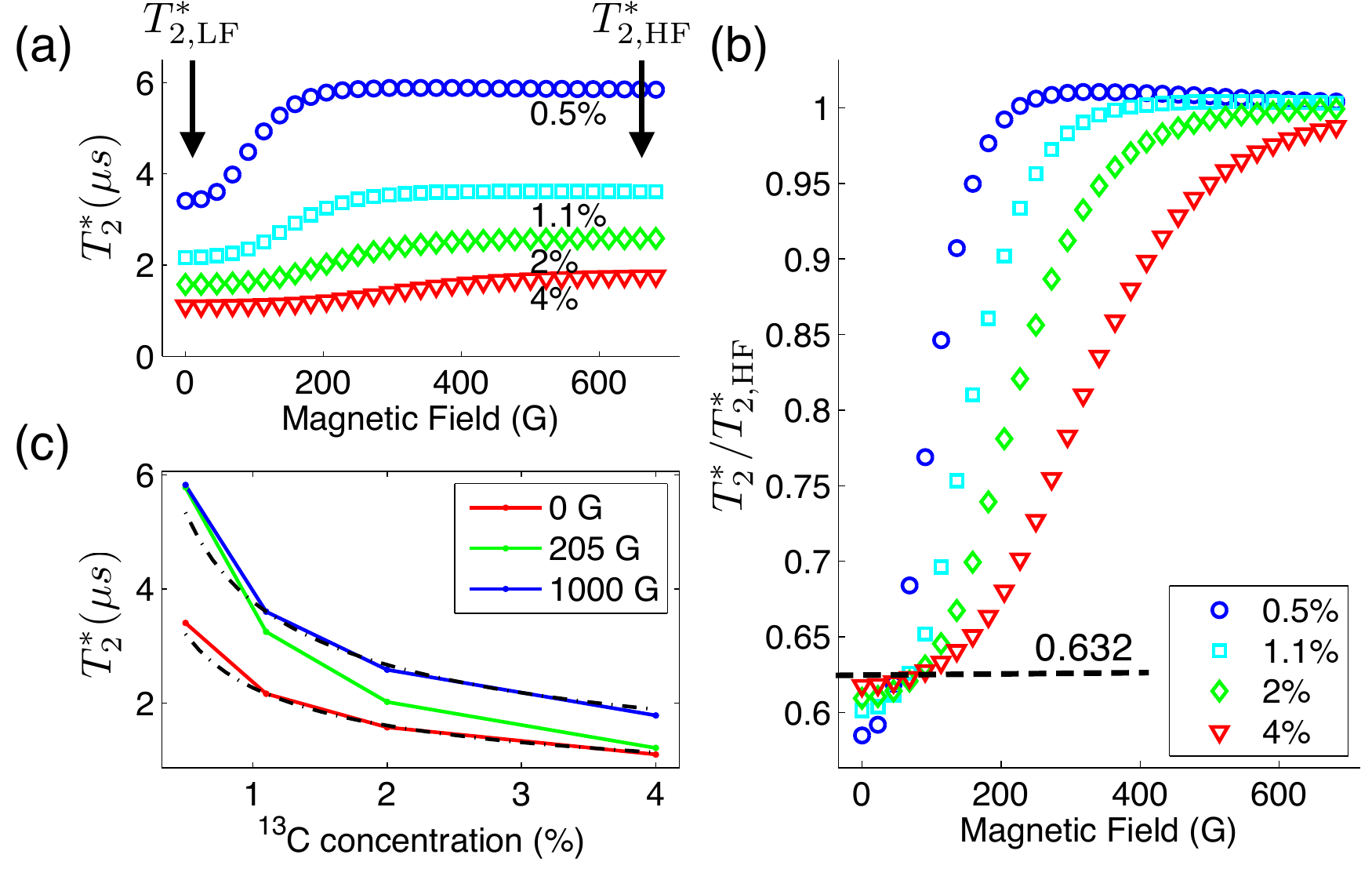}
\caption{(a) Coherence time as a function of the magnetic field strength for four different concentrations of $^{13}$C nuclear spin impurities. (b) Coherence time normalized to its value at high field ($T_{2,\rm{\tiny{ HF}}}^*$) as a function of the magnetic field magnitude for different concentrations of the nuclear spin bath. The enhanced factor is roughly $\sqrt{5/2} = 1/0.632$, independent of the concentration of impurities. (c) Coherence time as a function of the ${^{13}}$C concentration $p$. The simulations agree with the dash-dotted line proportional to $1/\sqrt{p}$ in the limit of either a weak or a high magnetic field. }
\label{fig5}
\end{center}
\end{figure*} 
We now study the dynamics of the FID signal as a function of the magnetic field amplitude for different concentrations of $^{13}$C nuclear spins $p$. For this purpose, we use the numerical method N2 which directly provides the average trend of the FID signal within short computational time. Figure~\ref{fig5}(a) shows the coherence time extracted from the envelope of the FID signal given by equation (\ref{eq:envelopep}), as a function of the magnetic field strength for four different concentrations of impurities. By decreasing the content of $^{13}$C, the coherence time of the central spin gets longer because $^{13}$C atoms are placed on average farther away from the central spin, and thus exhibit weaker hyperfine interaction~\cite{Balasubramanian:NatMat2009}. In addition, the magnetic field amplitude required to freeze the evolution due to the anisotropic component of the hyperfine interaction gets larger as the concentration of $^{13}$C nuclear spins increases. As a result, the magnetic field strength linked to the transition between the coherence time in the low field limit ($T_{2,\rm{\tiny{ LF}}}^*$) and in the high field limit $(T_{2,\rm{\tiny{HF}}}^*$) decreases with the $^{13}$C content. We note that for a given $^{13}$C content, the inner distribution of nearby nuclei might resemble to a bath configuration corresponding to smaller concentration. For this particular case, we expect to have a larger $T_2^*$ and a smaller magnetic field amplitude to increase $T_2^*$, as it might be the case for NV12 in figure \ref{fig3}(a).

The coherence time normalized to its value in the high field limit is shown in figure~\ref{fig5}(b). The enhancement of the coherence time matches very well the value given by equation~(\ref{enhance}), $T_{2,\rm{\tiny{ HF}}}^* / T_{2,\rm{\tiny{ LF}}}^* = \sqrt{5/2} = 1/0.632$, and is independent of the impurity concentration. Finally, the coherence time is plotted as a function of $p$ on figure~\ref{fig5}(c). In the limit of either a weak or a high magnetic field, the coherence time scales as $1/\sqrt{p}$, as expected from equations~(\ref{T2LF}) and~(\ref{T2HF}). In intermediate regimes, the scaling deviates from this simple behavior owing to the increment of the coherence time starting when $\gamma_{n}B\sim A^{(n)}$ (see figure \ref{fig5}(a)).

\subsection{Partition of the nuclear spin bath into shells and cones}\label{sec:shells}

In this section, we analyse the contribution to decoherence of each nuclear spin impurity with respect to its position relative to the central spin, using the numerical method N2 for a natural abundance of $^{13}$C.

\begin{figure*}[t]
\begin{center}
\includegraphics[width=1\textwidth]{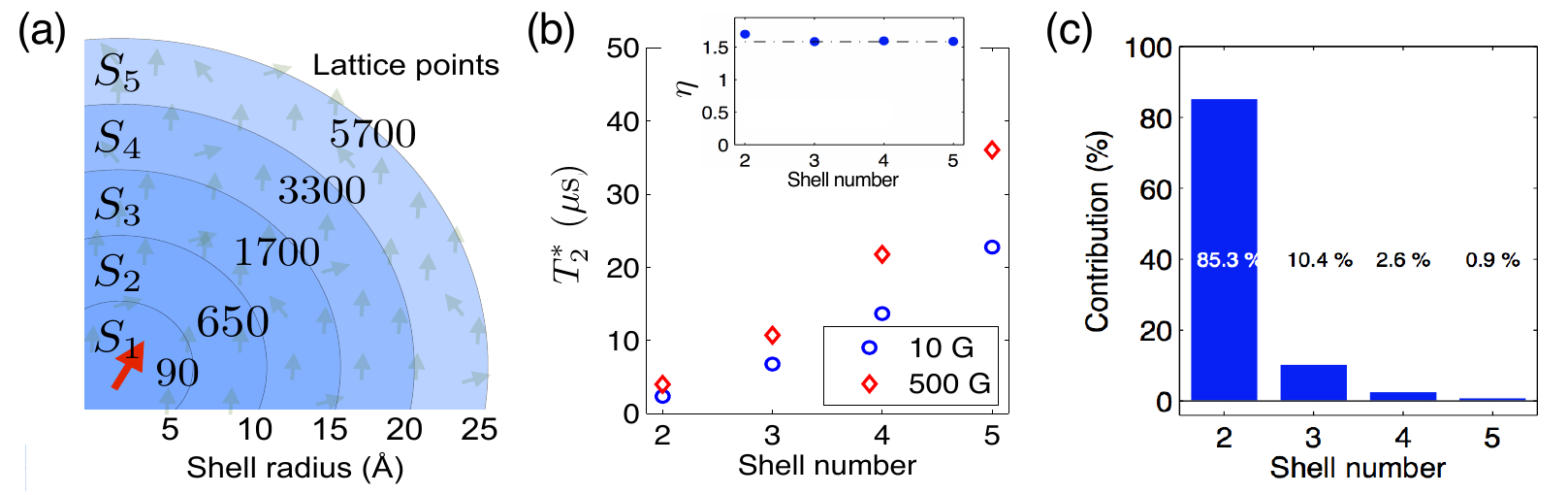}
\caption{(a) Shell partition of the nuclear spin bath. Each shell $S_{i}$ has a thickness of $5$~$\AA$. The number of lattice points per shell is indicated. (b) Coherence time versus shell number for low (10 G) and high (500 G) magnetic fields. The calculation is conducted with the numerical method N2 for a natural abundance of $^{13}$C. The coherence time enhancement for each shell is equal to $\sqrt{5/2}$, as shown in the inset. (c) Contribution of each shell to decoherence calculated as $(T_2^*/T_{2,S_{i}}^{*})^2$. As expected, nuclei close to the central spin contribute the most. }
\label{fig6}
\end{center}
\end{figure*}

First, we group ${^{13}}$C according to their radial distance to the central spin. The nuclear spin bath is thus divided into shells $S_{i}$ with a $5$ $\AA$ width. The number of lattice sites in each shell is roughly proportional to the area of the shell times the width, as shown in figure \ref{fig6}(a). To conduct the analysis, we infer the coherence time $T_{2,S_{i}}^*$ from the FID envelope due to the nuclei belonging to the shell $S_{i}$. 

For low and high magnetic fields, the total coherence time can be calculated as
\begin{eqnarray}
\left(\frac{1}{T_2^*}\right)^2 = \sum_i\left(\frac{1}{T_{2,S_{i}}^*}\right)^2 \ .
\end{eqnarray}
The coherence time for each shell at low and high magnetic fields is shown in figure \ref{fig6}(b). As expected, the closest shell to the central spin contributes the most to decoherence. We note that we have not considered strongly interacting nuclei, which results in neglecting the contribution of all lattice points of the first shell $S_{1}$, as explained in Section \ref{sec:num}. We estimate the contribution to decoherence of each shell as $\left( T_2^*/T_{2,S_{i}}^* \right)^2$. Figure \ref{fig6}(c) indicates that the second shell with $\approx 650$ lattice points, which corresponds to $\approx 7$ nuclear spins at $p_{nat}=1.1\%$, contributes in $85 \%$ to decoherence, meanwhile the third shell contributes only $10\%$. In addition, we note that, on average, the coherence time enhancement factor is not modified while changing the radial distance of the nuclear spins with respect to the central spin (see inset on figure \ref{fig6}(b)).\\

\begin{figure*}[t]
\begin{center}
\includegraphics[width=1.0\textwidth]{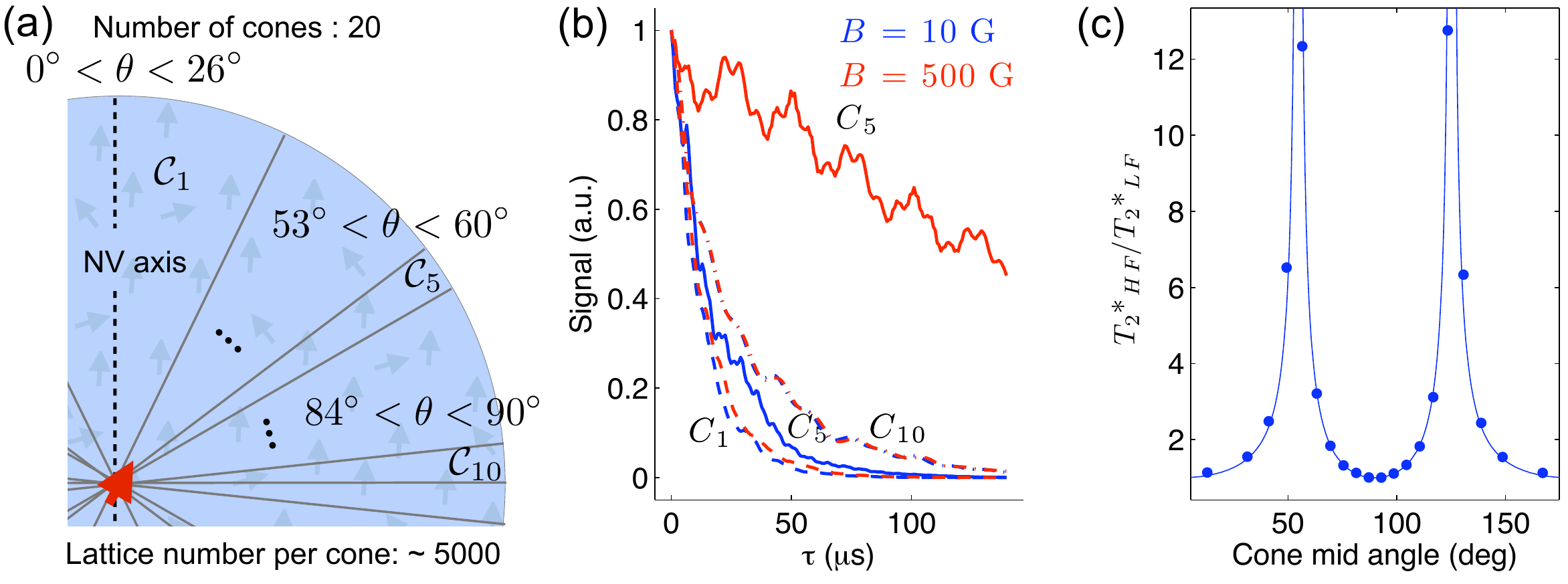}
\caption{(a) Cone partition of the nuclear spin bath. Each cone contains approximately the same number of lattice sites ($\sim 5000$). Cones $C_1$, $C_5$, and $C_{10}$ contain nuclear spin close to the north pole, to the magic angle $\theta_{M} \approx 54^\circ$, and to the equator, respectively. (b) Envelope of the FID signal for cones $C_1$ (dashed lines), $C_5$ (solid lines), and $C_{10}$ (dashed-dot lines) at weak magnetic field (blue) and at high magnetic field (red). (c) Coherence time enhancement factor $\eta$ as a function of the cone mid angle. The maximum enhancement occurs for nuclei with a polar angle $\theta$ close to the magic angle. The solid line is the theoretical prediction given by Equation~(\ref{final}).}
\label{fig:four}
\end{center}
\end{figure*}

We now analyze the contribution of each nuclear spin as a function of its angular orientation $\theta$ with respect to the symmetry axis of the NV defect. For this purpose, the nuclear spin bath is divided into 20 angular cones, as shown in figure \ref{fig:four}(a). Each cone $C_i$ contains on average an equal number of lattice sites, so that their relative contributions to decoherence can be better compared. The coherence time $T_{2,C_{i}}^*$ of the central spin is inferred by calculating the FID envelope due to the nuclei belonging to a given cone $C_i$ of the nuclear spin bath. 

Figure \ref{fig:four}(b) shows the FID envelope at weak and high magnetic fields for three different cones $C_1$, $C_5$, and $C_{10}$, containing nuclei close to the north pole, to the magic angle $\theta_{M} \approx 54^\circ$, and to the equator, respectively. For cones $C_1$ and $C_{10}$, the FID signals are almost identical at weak and high magnetic fields, whereas for cone $C_5$ there is a large enhancement of the coherence time at high magnetic field. The coherence time enhancement factor $\eta$ is plotted as a function of the cone mid angle in figure \ref{fig:four}(c). This behavior can be understood as follows. Within our model, the hyperfine interaction is assumed to be purely dipolar between the central spin and each nuclear spin of the bath. In this framework, equations~(\ref{eq:A}) and (\ref{eq:B}) indicate that the anisotropic component of the hyperfine vector $A_{\perp}$ is zero at the poles ($\theta = \kp{0,\pi}$) and at the equator ($\theta=\pi/2$), meanwhile the $A_{zz}$ component is zero at the polar magic angle ($\theta_{M} = \cos^{-1}(1/\sqrt{3})\approx 54^\circ$). Therefore, since the anisotropic component is suppressed for large magnetic fields, the main contribution to the coherence time enhancement comes from the nuclei with axial component $A_{zz}\approx0$, {\it i.e.} placed at the magic angle $\theta_{M} \approx 54^\circ$. This effect is an analogous phenomenon to magic angle spinning in solid-state nuclear magnetic resonance spectroscopy~\cite{Slichter:1996}. Following equation (\ref{eq:enhancement}), the enhancement factor as a function of the polar angle $\theta$ can be written as
\begin{eqnarray}
\eta(\theta) = \frac{T_{2,\rm{\tiny{HF}}}^*(\theta)}{T_{2,\rm{\tiny{ LF}}}^*(\theta)} = \left| \frac{\sqrt{1+3\cos^2\theta}}{1-3\cos^2\theta} \right| \ .
\label{final}
\end{eqnarray}
This formula accurately matches the enhancement factor inferred from the numerical simulations, as shown in figure~\ref{fig:four}(c).
 
From the partition of the bath into shells and cones, we conclude that nuclear spin baths with a large number of nuclei close to the magic angle and close to the central spin exhibit a large enhancement of their coherence time when the magnetic field is increased. This feature also explains the large variation of the enhancement factor shown in figures~\ref{fig4}(c) and (d), which correlates with the presence of nuclei close to the magic cone.

\section{Conclusions}
We have studied the decoherence of the electronic spin associated with a single NV defect in diamond placed in a $^{13}$C nuclear spin bath. By recording the FID signal as a function of the strength of a magnetic field applied along the NV defect axis, we have demonstrated an increment of the coherence time at high magnetic field, in agreement with theoretical studies. Numerical simulations of the FID signal indicate that the coherence time enhancement is independent of the $^{13}$C concentration and is mainly due to nuclei positioned close to the vacancy and close to the polar magic angle ($\theta\approx 54^\circ$) with respect to the symmetry axis of the NV defect. This work points out interesting dynamics of a single spin placed in a spin bath, whose understanding is required for the development of diamond-based quantum information processing where single spins in diamond are used as robust solid-state qubits.

\ack
The authors acknowledge L.~Childress, P.~Bertet, R.~Hanson, P.~Spinicelli, M.~Lesik and M. Lecrivain for fruitful discussions and experimental assistance. This work is supported by the Agence Nationale de la Recherche (ANR) through the projects A{\sc dvice} and Q{\sc invc}, from Conicyt Fondecyt, Grant No. 11100265, Millennium Scientific Initiative P10-035-F, and US Air Force Grant FA9550-12-1-0214.

\appendix
\section{}
In this appendix, we show how to obtain the expressions for $\mathcal{S}_n(\tau)$ and the envelope $\bar{E}(\tau)$ in the low and high magnetic field limits (equations (\ref{eq:psu}),(\ref{ELF}) and (\ref{EHF})). We start from equation (\ref{eq:s}) and consider,
\begin{eqnarray}
\mathcal{S}_n(\tau) = \textrm{Tr}_n\left\{  \rho_n U_1^{(n)\dagger} U_0^{(n)}  \right\}
\end{eqnarray}
where
\begin{eqnarray}
U_{m_s}(\tau) = \exp{[-i(\mathbf{\Omega}_{m_s} \cdot \mathbf{I}})\tau] = \cos \frac{\Omega_{m_s}\tau}{2} -i(\mbox{\boldmath{$\sigma$}}\cdot \mathbf{u}_{m_s})\sin \frac{\Omega_{m_s}\tau}{2} \ .
\end{eqnarray}
Here {\boldmath{$\sigma$}} is the vector of Pauli spin matrices and $\mathbf{u}_{m_s}=\mathbf{\Omega}_{m_s}/|\mathbf{\Omega}_{m_s}|$. The superscript $n$ has been omitted for simplicity. Consider
\begin{eqnarray}
U_1^\dagger U_0 = \cos \frac{\Omega_0\tau}{2}\cos \frac{\Omega_1\tau}{2} -i(\mbox{\boldmath{$\sigma$}}\cdot \mathbf{u}_{0})\cos \frac{\Omega_1\tau}{2}\sin \frac{\Omega_0\tau}{2} \\ -i(\mbox{\boldmath{$\sigma$}}\cdot \mathbf{u}_{1})\cos \frac{\Omega_0\tau}{2}\sin \frac{\Omega_1\tau}{2} + (\mbox{\boldmath{$\sigma$}}\cdot \mathbf{u}_{0})(\mbox{\boldmath{$\sigma$}}\cdot \mathbf{u}_{1}) \sin \frac{\Omega_0\tau}{2}\sin \frac{\Omega_1\tau}{2}
\end{eqnarray}
where $(\mbox{\boldmath{$\sigma$}}\cdot \mathbf{u}_{0})(\mbox{\boldmath{$\sigma$}}\cdot \mathbf{u}_{1}) = \mathbf{u}_{0}\cdot \mathbf{u}_{1} + i \mbox{\boldmath{$\sigma$}}\cdot \mathbf{u}_{0} \times \mathbf{u}_{1}$. For $\rho_n= \mathbbm{1}/2$, when the trace is taken over the Hilbert space of nucleus $n$, only those terms not proportional to $\sigma$ will survive because $\textrm{Tr}_n(\mbox{\boldmath{$\sigma$}}\cdot \mathbf{u}_{m_s})=0$. Therefore,
\begin{eqnarray}\label{eq:psn}
\mathcal{S}_n(\tau) = \cos \frac{\Omega_0^{(n)}\tau}{2}\cos \frac{\Omega_1^{(n)}\tau}{2} + \left[\mathbf{u}_{0}^{(n)}\cdot \mathbf{u}_{1}^{(n)} \right]\sin \frac{\Omega_0^{(n)}\tau}{2}\sin \frac{\Omega_1^{(n)}\tau}{2}
\end{eqnarray}
where we identify $\mathbf{u}_{0}^{(n)}\cdot \mathbf{u}_{1}^{(n)} = \cos\beta^{(n)}$ as the angle between the two Larmor vectors. Note that in our construction  $\Omega_0^{(n)} = \gamma_nB$, for all nuclei. Equation (\ref{eq:psn}) can be arranged as
\begin{eqnarray}\label{eq:psn2}
\mathcal{S}_n(\tau) = \cos \frac{(\Omega_0^{(n)}-\Omega_1^{(n)})\tau}{2} + (\cos\beta^{(n)} - 1)\sin \frac{\Omega_0^{(n)}\tau}{2}\sin \frac{\Omega_1^{(n)}\tau}{2},
\end{eqnarray}
where the second term is zero for both the low and high field limits and it can be considered as the transition function between the two regimes. In the low field limit $B\approx 0$,
\begin{eqnarray}
\mathcal{S}_n(\tau) \approx \cos \frac{\Omega_1^{(n)}\tau}{2} = 1 - \frac{1}{2}\left(\frac{\Omega_1^{(n)}\tau}{2}\right)^2 = 1 - \frac{1}{8}[ A^{(n)} \tau]^2.
\end{eqnarray}
From equation (\ref{eq:envelopep}), the envelope of the FID signal is,
\begin{eqnarray}
\bar{E}(\tau) \approx  \textrm{Re} \Pi_n \left[ p\left( 1 - \frac{1}{8}[ A^{(n)} \tau]^2 \right) + 1 - p  \right] = 
\Pi_n \left( 1- \frac{p}{8}[ A^{(n)} \tau]^2 \right).
\end{eqnarray}
In the high field limit ($\gamma_nB\gg A^{(n)}$), $\cos\beta^{(n)}\sim 1$, and 
\begin{eqnarray}
\mathcal{S}_n(\tau) \approx\cos \frac{(\Omega_0^{(n)}-\Omega_1^{(n)})\tau}{2} \approx 1 - \frac{1}{8}[A_{zz}^{(n)}\tau]^2
\end{eqnarray}
where we used $\Omega_1^{(n)} = \left( A_\perp^{(n)2} + (\gamma_nB+A_{zz}^{(n)})^2 \right)^{1/2}\approx \gamma_n B \left( 1 + \frac{2A_{zz}^{(n)}}{\gamma_nB} + \left(\frac{A^{(n)}}{\gamma_nB}\right)^2 \right)^{1/2}$ $\approx \gamma_n B + A_{zz}^{(n)}$. Therefore, the envelope signal in this limit is
\begin{eqnarray}
\bar{E}(\tau) \approx  \textrm{Re} \Pi_n \left[ p\left( 1 - \frac{1}{8}[ A_{zz}^{(n)} \tau]^2 \right) + 1 - p  \right] = 
\Pi_n \left( 1- \frac{p}{8}[ A_{zz}^{(n)} \tau]^2 \right).
\end{eqnarray}

\section*{References}
\bibliographystyle{iopart-num}	
\bibliography{References}

\end{document}